\documentclass[conference]{IEEEtran}
\usepackage{amsfonts}
\usepackage{amssymb}

\usepackage{cite}

\ifCLASSINFOpdf
\usepackage[pdftex]{graphicx}
\graphicspath{{./Images/}}
\else
\fi
\usepackage[cmex10]{amsmath}
\usepackage{algorithmic}
\usepackage{algorithm}

\usepackage{array}

\usepackage[tight,footnotesize]{subfigure}
\DeclareMathOperator*{\argmin}{arg\,min}
\IEEEoverridecommandlockouts
\begin{document}
%
\title{SPRT Based Transceiver for Molecular Communications \vspace{-1.0cm}
\thanks{This research was funded in part by NSF grants CCF-1718560 and CCF-1817200.}
}

\author{Tze-Yang Tung and Urbashi Mitra \\
University of Southern California\\
\{tzeyangt,ubli\}@usc.edu
}

\maketitle

\begin{abstract}

Achieving precise synchronisation between transmitters and receivers is particularly challenging in diffusive molecular communication environments. To this end, point-to-point molecular communication system design is examined wherein synchronisation errors are explicitly considered. Two transceiver design questions are considered: the development of a sequential probability ratio test-based detector which allows for additional observations in the presence of uncertainty due to mis-synchronisation at the receiver, and a modulation design which is optimised for this receiver strategy. The modulation is based on optimising an approximation for the probability of error for the detection strategy and directly exploits the structure of the probability of molecules hitting a receiver within a particular time slot. The proposed receiver and modulation designs achieve strongly improved asynchronous detection performance for the same data rate as a decision feedback based receiver by a factor of 1/2.

\end{abstract}



\vspace{-0.6cm}

\section{Introduction}
\label{sec:intro}

Nano-machines are proposed to enable future medical and biological applications such as precision drug delivery and immune system support \cite{nakano_molecular_2012}. A necessary element in such systems is point-to-point molecular communication; a suggested strategy is communication via molecular diffusion. A nano-machine transmitter is assumed to contain a storage of molecules that are released into the medium. Diffusion carries the molecules across the communication channel, which are then detected by the receiver nano-machine. Different modulation schemes have been proposed for diffusive molecular communication \cite{kuran_modulation_2011,arjmandi_diffusion-based_2013}; however, all such schemes suffer from extensive 
inter-symbol interference (ISI) due to the nature of diffusive channels. 

Early molecular communication works presumed perfect timing information (synchronisation) at the receiver \cite{mosayebi_receivers_2014,movahednasab_adaptive_2016,zare_receiver_2017}. To address this issue, a variety of synchronisation schemes have been proposed \cite{hsu_training-based_2017,jamali_symbol_2017,lin_clock_2016,shahmohammadian_blind_2013,lee_asynchronous_2015}. However, the absolute performances of these methods are limited in diffusive channels and are often high in complexity, making them prohibitive for nano-machines. Recognising that precise synchronisation is challenging to achieve, a simple asynchronous detection scheme is introduced in \cite{noel_asynchronous_2017}. However, as will be shown, the performance is not always strong. We will show that with a moderate increase in complexity, performance can be improved drastically.


Herein, we presume the linear time-invariant (LTI) Poisson channel of \cite{arjmandi_diffusion-based_2013}, based on the additive inverse Gaussian channel model of \cite{srinivas_molecular_2012}. We develop the {\em Memory Aided Sequential Probability Ratio Test} (MASPRT).
MASPRT is an augmented version of the sequential probability ratio test (SPRT) \cite{wald_sequential_1945}.  The effect of ISI is mitigated through the feedback of prior decisions. The MASPRT offers resilience to synchronisation errors by limiting the effect of likelihood function mismatch due to the synchronisation errors.  Additionally,  the modulation is optimised for use in the MASPRT.  The proposed scheme is compared to that in  \cite{mosayebi_receivers_2014} which was shown to be near-optimal for the LTI Poisson channel and under the assumption of perfect synchronisation; as well as the detector proposed in  \cite{noel_asynchronous_2017}  which was designed to be resilient to timing errors. The MASPRT offers superior performance to both schemes in the presence of mis-synchronisation.

The main contributions of this paper are as follows; (i) a novel transceiver for molecular communication based on SPRT, called MASPRT is proposed, which utilises decision feedback and random stopping times to tolerate greater synchronisation errors compared to other asynchronous detection schemes; (ii) bounds on the error probability and expected stopping time of MASPRT are calculated; (iii) the bounded performance metrics are used to pose a modulation optimisation problem; and finally the performance of the overall scheme is compared to prior art.

 This paper is organised as follows: Section \ref{sec:model} and \ref{sec:prop_rx} describe the channel and transceiver models, respectively; Section \ref{sec:opt} proposes the optimal transceiver design and Section \ref{sec:results} discusses the numerical results based on bounding key performance metrics in the proposed design. 

\vspace{-0.1cm}

\section{Received Signal}
\label{sec:model}

\begin{figure}
        \centering
        \includegraphics[width=0.9\linewidth]{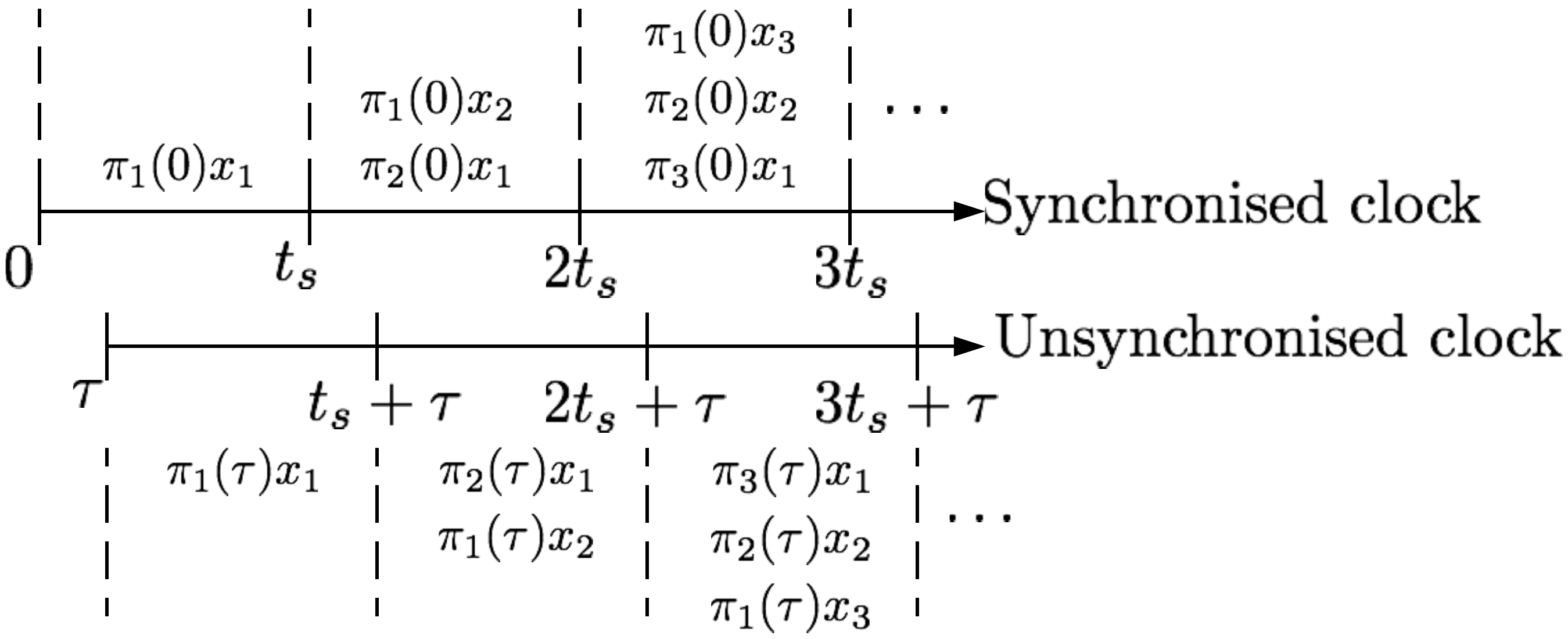}
        \caption{Illustration of synchronisation error on sample distribution.}
        \label{fig:sync_error}
        \vspace{-0.5cm}
\end{figure}

We assume diffusion-based molecular communication where molecules experience Brownian motion without drift. Without loss of generality, we consider one-dimensional motion. Molecules are released from a storage where  the number of released molecules is a Poisson random variable with rate $x$. This distribution is indicated by $\mbox{Poi}(x)$. The sequence of rates employed constitutes the transmitted message. The received signal is described by the one-dimensional linear-time invariant Poisson model proposed in\cite{arjmandi_diffusion-based_2013}. We assume a constant sampling duration denoted by $t_s$ and the rate $x_i$ denotes the rate associated with time slot $i$. Assuming that the molecules are released at the beginning of each time slot, the probability that a molecule is detected by the receiver at the $i^{\text{th}}$ time slot is denoted by $\pi_i(\tau)$, where $\tau$ is the synchronisation error between the transmitter and receiver clocks as illustrated in Figure \ref{fig:sync_error}. $\pi_i(\tau)$ is defined as\cite{arjmandi_diffusion-based_2013}
\begin{equation}
        \pi_i(\tau) =2Q\bigg(\frac{\rho}{\sqrt{it_s + \tau}}\bigg)-
        2Q\bigg(\frac{\rho}{\sqrt{(i-1)t_s + \tau}}\bigg),
        \label{eq:p_k}
\end{equation}
where the constant $\rho\triangleq\frac{d}{\sqrt{2D}}$ summarises the relationship between
the diffusion constant ($D$) and the distance between the transmitter and
receiver ($d$). The standard Q function is given by $Q(a)=\int_{a}^\infty\frac{1}{\sqrt{2\pi}}e^{\frac{-x^2}{2}}dx$. The noise within the system is modelled as a homogeneous Poisson process, with rate function $\lambda(t)=\lambda_0t$. Inter-molecular collisions are assumed to have negligible effects on the diffusion of molecules. As such, the number of molecules detected within the $i^{\text{th}}$ time slot is a Poisson random variable with the following distribution
\begin{equation}
        Y_i\sim\mbox{Poi}(\pmb{\pi}_i(\tau) \otimes \pmb{x}_i+\lambda_0t_s), \quad \forall i\geq1,
        \label{eq:y_i}
\end{equation}
where $\pmb{\pi}_i(\tau)=[\pi_1(\tau),\pi_2(\tau),\dots,\pi_i(\tau)]$, $\pmb{x}_i=[x_1,x_2,\dots,x_i]$ and $\otimes$ denotes the convolution operation. The samples are independent. 





We shall assume binary signalling wherein the source symbol, $s_j$, maps to a modulation signal denoted by $\mathbf{x}_j=[x_{1|j},x_{2|j},\dots,x_{N|j}] \in \mathbb{R}^N$, $j=\{0,1\}$. 
In the sequel, we will optimise the designs of $\mathbf{x}_j$ for detection performance and data rate. The transmission rate is defined as $R=\frac{1}{t_s\cdot N}$. It is assumed that $||\mathbf{x}_j||_2\leq P$, where $P$ is the power constraint.

\vspace{-0.1cm}

\section{Memory Aided SPRT}
\label{sec:prop_rx}

To increase robustness to synchronisation errors, we consider detection strategies based on sequential probability ratio tests (SPRT) \cite{wald_sequential_1945}. As the original SPRT presumes a one time decision with random stopping time, we adapt this scheme and compensate for ISI using memory, hence the {\em Memory Aided Sequential Probability Ratio Test} (MASPRT). Such schemes terminate with a decision only when confident; if there is uncertainty, additional samples are taken. Let $\mathbf{y}_m=[y_1,y_2,\dots,y_m]$, the likelihood ratio function is given by
\begin{eqnarray*}
L_m(\mathbf{y}_m) & = & \frac{{p}_{\mathbf{y}_m|s_1}}{{p}_{\mathbf{y}_m|s_0}},
\end{eqnarray*}
where ${p}_{\mathbf{y}_m|s_j}$ is the joint Poisson probability mass function for symbol $s_j$ as implied by Equation (\ref{eq:y_i}).  If $m <N$, for an $N$ to be described in the sequel, the decision rule is given by
\begin{eqnarray*}
\delta(L_m(\mathbf{y}_m)) & = & \left\{ \begin{array}{l l}
s_0, & L_m(\mathbf{y}_m) \leq A\\
s_1, &L_m(\mathbf{y}_m) \geq B\\
\mbox{sample}, & \mbox{else}
\end{array} 
\right.
\label{eq:decision}
\end{eqnarray*}
The constants $A$ and $B$ are given by $A=\frac{\beta}{1-\alpha}, \; B=\frac{1-\beta}{\alpha}$ (as in \cite{wald_sequential_1945}),
where $\alpha$ is the false alarm rate and $\beta$ is the miss probability. If the MASPRT does not terminate by $m=N$, a truncation rule is applied to ensure that one does not sample into the subsequent channel symbol:
\begin{eqnarray*}
\delta(L_N(\mathbf{y}_N)) & = & \left\{ \begin{array}{l l}
s_0, & |L_N(\mathbf{y}_N) - A| < |L_N(\mathbf{y}_N) - B| \\    
s_1,& |L_N(\mathbf{y}_N) - A| \geq |L_N(\mathbf{y}_N) - B|\\
\end{array}.
\right.
\label{eq:trunc}
\end{eqnarray*}



The likelihood ratio $L_m(\mathbf{y}_m)$ requires the knowledge of the joint likelihood function of the samples $p_{\mathbf{y}_m|s_j}$ for $j\in\{0,1\}$. However, since the channel exhibits memory of all past transmissions, knowledge of all past diffusion rates would be needed to calculate the exact likelihood function, which is infeasible. To remedy this issue, a memory of $B$ bits is embedded in the receiver, such that the receiver can store $B$ past decoded source symbols at each time slot $i$, denoted by $M_i=\{\hat{s}_{i-1},\hat{s}_{i-2},\dots,\hat{s}_{i-B}\}$. The receiver is then able to estimate the likelihood function $\hat{p}_{\mathbf{y}_m|s_j}$ up to $B\cdot N+m$ time slots in the past by appropriately using the mapping $s_j\rightarrow\mathbf{x}_j$. Assuming the source symbols in the memory were decoded correctly, the more memory the receiver possesses, the more accurate the likelihood ratio and the lower the error probability for MASPRT. It should be noted here that the desired $\alpha$ and $\beta$ can only be achieved if the likelihood ratio is calculated exactly. Since an estimate is used here, the probability of error will be greater than those set by $\alpha$ and $\beta$, depending on how much memory is used and the synchronisation error. We emphasise that the receiver assumes that $\tau=0$, which results in errors in the calculation of the likelihood ratio in the presence of synchronisation errors.





\section{Transceiver Optimisation}
\label{sec:opt}

In \cite{mosayebi_receivers_2014}, it is shown that the optimal value for $\mathbf{x}_0$ is $\mathbf{x}_0=\mathbf{0}$ (vector of zeros) for the maximum-likelihood decision aided (MLDA) receiver. We adopt this result and optimise $\mathbf{x}_1$. The error probability is given by $P_e=P[s_0]P[s_1|s_0]+P[s_1]P[s_0|s_1]$. Given Wald's threshold approximations \cite{wald_sequential_1945}, we assume that $P[s_1|s_0]\!\!>>\!\!P[s_0|s_1]$ due to ISI of past $s_1$ transmissions. Hence, we focus on the minimisation of $P[s_1|s_0]$. Bounds on the expected stopping times under $s_0$ and $s_1$ are also derived and the value of $N$ selected such that the probability of engaging the truncation rule is small.

%
\subsection{Bounding  $P[s_1|s_0]$}
\label{subsec:bound_Pe}

The analysis of the error probability of SPRT for non-independent and identically distributed samples is very challenging. Thus, we provide the expression for $P[s_1|s_0]$ and apply bounding techniques to yield an expression that is amenable to optimisation. 

\noindent
\emph{Proposition 1}: We assume an LTI Poisson channel as described in Equation (\ref{eq:y_i}) and that $\mathbf{x}_0=\mathbf{0}$. Let $E[Y_k|s_j]=\bar\lambda_{k|j}(\mathbf{x}_1)=\bar\lambda_{k}^C+\bar\lambda_{k}^{ISI}+n_0$, where $\bar\lambda_{k|j}^C$ denotes the diffusion rates transmitted for the current source symbol, $\bar\lambda_{k}^{ISI}$ denotes the ISI terms, and $n_0=\lambda_0t_s$. Then $P[s_1|s_0]$ can be bounded as follows:
\begin{align*}
        &P[s_1|s_0]\leq\bar\lambda_{1|1}^C+\mu\sum_{k=2}^N\bar\lambda_{k|1}^C\\
        &=\bigg(\pi_1(\tau)+\mu\sum_{i=2}^N\pi_i(\tau)\bigg)x_{1|1}\!+\!\mu\sum_{k=1}^{N-1}x_{k+1|1}\sum_{i=1}^{N-k}\pi_i(\tau).
\end{align*}
where $\mu=P\Bigg[\frac{\log(A)+2\pi_1(\tau)x_{1|1}}{\log\big(\frac{x_{1|1}+\lambda_1^{ISI}+n_0}{n_0+\lambda_1^{ISI}}\big)}<y_1<\frac{\log(B)+2\pi_1(\tau)x_{1|1}}{\log\big(\frac{x_{1|1}+\lambda_1^{ISI}+n_0}{n_0+\lambda_1^{ISI}}\big)}\Bigg]$.

\noindent
\emph{Sketch of Proof}: Consider the joint log-likelihood ratio (LLR) up to time $T$, where $T$ is an integer valued random variable with $E[T]<\infty$:
\begin{align*}
        \log[L_T(\mathbf{y}_T)] &=\sum_{k=1}^T\log\frac{{p}_{{y}_k|s_1}}{{p}_{{y}_k|s_0}}\\
        &=\sum_{k=1}^Ty_k\log\bigg(\frac{\bar\lambda_{k|1}^{C}+\bar\lambda_k^{ISI}+n_0}{n_0+\bar\lambda_k^{ISI}}\bigg)-\bar\lambda_{k|1}^{C}\\
        &=\sum_{k=1}^T\alpha_ky_k-g_T(\mathbf{x}_1),
\end{align*} 
where 
\begin{align}
        \alpha_k&=\log\bigg(\frac{\bar\lambda_{k|1}^{C}+\bar\lambda_k^{ISI}+n_0}{n_0+\bar\lambda_k^{ISI}}\bigg),\\
        g_T(\mathbf{x}_1)&=\sum_{k=0}^{T-1}x_{T-k|1}\sum_{i=1}^{k+1}\pi_i(\tau).
\end{align}

Let $c_k^B=\log(B)+g_k(\mathbf{x}_1)$, $c_k^A=\log(A)+g_k(\mathbf{x}_1)$, $\pmb{\alpha}_k=[\alpha_1,\alpha_2,\dots,\alpha_k]$, $\mathbf{y}_k=[y_1,y_2,\dots,y_k]$ and the event $\{s_1|s_0\}_{T=t}$ denotes an error at time $T=t$. Given that the sampling limit is $N$ and neglecting the error probability of the truncation rule, $P[s_1|s_0]$ can be written as:
\begin{align*}
        &~~~~P[s_1|s_0]\\
        &=E_T\big\{P[\{s_1|s_0\}_{T=t}\big|T=t]\}\\
        &\approx\!\sum_{t=1}^N\!P[\pmb{\alpha}_t^T\mathbf{y}_t\geq c_t^B,\!\cap_{k=1}^{t-1}(c_k^A<\!\pmb{\alpha}_k^T\mathbf{y}_k\!<c_k^B)|T=t]P[T=t]\\
        &\stackrel{(a)}{=}\sum_{t=1}^NP[\pmb{\alpha}_t^T\mathbf{y}_t\geq c_t^B,\cap_{k=1}^{t-1}(c_k^A<\pmb{\alpha}_k^T\mathbf{y}_k<c_k^B)]\\ 
        &\stackrel{(b)}{\leq}\! P\Big[y_1\geq\frac{c_1^B}{\alpha_1}\Big]\!+\!P[c_1^A<\!\alpha_1y_1\!<c_1^B]\sum_{k=2}^N\!P\Big[y_k\geq \frac{c_k^B\!-\!c_{k-1}^B}{\alpha_k}\Big]\!\\
        &\stackrel{(c)}{\leq}\frac{\bar\lambda_{1|0}(\mathbf{x}_1)\alpha_1}{c_1^B}+\mu\sum_{k=2}^N\frac{\bar\lambda_{k|0}(\mathbf{x}_1)\alpha_k}{c_k^B-c_{k-1}^B}\\
        &\stackrel{(d)}{\leq}\frac{\bar\lambda_{1|0}(\mathbf{x}_1)\frac{\bar\lambda_{1|1}^C}{n_1+\bar\lambda_1^{ISI}}}{\log(B)+g_1(\mathbf{x}_1)}+\mu\sum_{k=2}^N\frac{\bar\lambda_{k|0}(\mathbf{x}_1)\frac{\bar\lambda_{k|1}^C}{n_k+\bar\lambda_k^{ISI}}}{g_k(\mathbf{x}_1)-g_{k-1}(\mathbf{x}_1)}\\ 
        &\stackrel{(e)}{=}\frac{\bar\lambda_{1|1}^C}{\log(B)+g_1(\mathbf{x}_1)}+\mu\sum_{k=2}^N\frac{\bar\lambda_{k|1}^C}{g_k(\mathbf{x}_1)-g_{k-1}(\mathbf{x}_1)}\\
        &\stackrel{(f)}{\leq}\bar\lambda_{1|1}^C+\mu\sum_{k=2}^N\bar\lambda_{k|1}^C.
\end{align*}
The error probability of the truncation rule is neglected as the value of $N$ will be chosen such that the probability of engaging the rule is small. In Step $(a)$, the event $\{T=t\}=\{\pmb{\alpha}_t^T\mathbf{y}_t\geq c_t^B,\cap_{k=1}^{t-1} (c_{k}^A<\pmb{\alpha}_{k}^T\mathbf{y}_{k}<c_{k}^B)\}\cup\{\pmb{\alpha}_t^T\mathbf{y}_t\leq c_t^A,\cap_{k=1}^{t-1} (c_{k}^A<\pmb{\alpha}_{k}^T\mathbf{y}_{k}<c_{k}^B)\}$, which yields the result. Step $(b)$ follows from taking the upper bound of $\pmb{\alpha}_{m-1}^T\mathbf{y}_{m-1}$ to lower bound the value of $\alpha_my_m$ and that the samples are independent. Inequalities $(c)$ and $(d)$ arise from using the Markov Inequality\cite{leon-garcia_probability_2008} and $x>\log(1+x)$, respectively. Equality $(e)$ follows from the definition of $\bar\lambda_{k|0}$ and the final inequality, $(f)$, follows from assuming the denominators in equality $(e)$ are $\geq1$. It should be noted here that the upper bound for $P[s_1|s_0]$ derived herein is quite loose, but as will be seen in Section \ref{sec:results}, the result of optimising $\mathbf{x}_1$ using this bound yields strong performance improvements over non-optimised signals.

\subsection{Bounding expected stopping time}
\label{subsec:bound_Estopp}
In order to define our optimisation problem for $\mathbf{x}_1$, we need bounds on the expected stopping times of the MASPRT. 

\noindent
\emph{Proposition 2}: There exists $\epsilon>0$ and $T_j(\epsilon)$ for $j\in\{0,1\}$, such that:
\begin{align*}
        \!\begin{cases}
        \!\sum\limits_{k=T_1(\epsilon)+1}^{\infty}\frac{1}{k}\bigg(\!\bar\lambda_{k|1}\!(\!\mathbf{x}_1\!)\!\log\!\bigg(\!\frac{\bar\lambda_{k|1}(\mathbf{x}_1)}{\bar\lambda_{k|0}(\mathbf{x}_1)}\!\bigg)\!+\!\bar\lambda_{k|0}(\!\mathbf{x}_1\!)\!-\!\bar\lambda_{k|1}\!(\!\mathbf{x}_1\!)\!\bigg)\!\leq\epsilon\\
        \!\sum\limits_{k=T_0(\epsilon)+1}^{\infty}\frac{1}{k}\bigg(\!\bar\lambda_{k|0}\!(\!\mathbf{x}_1\!)\!\log\!\bigg(\!\frac{\bar\lambda_{k|0}(\mathbf{x}_1)}{\bar\lambda_{k|1}(\mathbf{x}_1)}\!\bigg)\!+\!\bar\lambda_{k|1}(\!\mathbf{x}_1\!)\!-\!\bar\lambda_{k|0}\!(\!\mathbf{x}_1\!)\!\bigg)\!\leq\epsilon,\\
        \end{cases}
\end{align*}
and
\begin{align*}
        \!\begin{cases}
        \!\frac{\log B}{T_1(\epsilon)}\leq\sum\limits_{k=1}^{T_1(\epsilon)}\frac{1}{k}\bigg(\!\bar\lambda_{k|1}\!(\!\mathbf{x}_1\!)\!\log\!\bigg(\!\frac{\bar\lambda_{k|1}(\mathbf{x}_1)}{\bar\lambda_{k|0}(\mathbf{x}_1)}\!\bigg)\!+\!\bar\lambda_{k|0}(\!\mathbf{x}_1\!)\!-\!\bar\lambda_{k|1}\!(\!\mathbf{x}_1\!)\!\bigg)\!\\
        \!\frac{-\log A}{T_0(\epsilon)}\leq\sum\limits_{k=1}^{T_0(\epsilon)}\frac{1}{k}\bigg(\!\bar\lambda_{k|0}\!(\!\mathbf{x}_1\!)\!\log\!\bigg(\!\frac{\bar\lambda_{k|0}(\mathbf{x}_1)}{\bar\lambda_{k|1}(\mathbf{x}_1)}\!\bigg)\!+\!\bar\lambda_{k|1}(\!\mathbf{x}_1\!)\!-\!\bar\lambda_{k|0}\!(\!\mathbf{x}_1\!)\!\bigg)\!.
        \end{cases}
\end{align*}
\emph{Sketch of Proof}:
We first provide a needed lemma,

\noindent
\emph{Lemma 1}\cite{wald_cumulative_1944}: Let $\{Z_t\}_{t\in\mathbb{N}}$ be a sequence of real valued random variables and $T$ be non-negative integer valued random variable. If $\{Z_t\}_{t\in\mathbb{N}}$ is integrable $\forall t$, $E[T]<\infty$, $\exists~\sigma$ such that $E[|Z_t|]\leq\sigma~\forall t$, a filtration\cite{leon-garcia_probability_2008} $\{\mathcal{F}_t\}_{t\in\mathbb{N}_0}$ such that $T$ is a stopping time for the filtration, and $Z_t$ is independent of $\mathcal{F}_{t-1}~\forall t$, then the event $\{T\geq t\}=\{T\leq t-1\}^c\in\mathcal{F}_{t-1}$ and is independent of $Z_t$.

Let $\mathbb{I}_{\{\cdot\}}$ be the indicator function, $E_j[T]$ be the expected stopping time under symbol $s_j$, $j\in\{0,1\}$, and $D(p_a||p_b)=E_a[\log\frac{p_a}{p_b}]$ be the Kullback-Leibler (K-L) distance between distributions $p_a$  and $p_b$. Then:
\begin{align*}
        &E_j[\log(L_T(\mathbf{y}_T))]=E_j\bigg[\sum_{k=1}^T\log\frac{{p}_{{y}_k|s_1}}{{p}_{{y}_k|s_0}}\bigg]\\
        &=E_j\bigg[\sum_{k=1}^\infty\mathbb{I}_{T\geq k}\log\frac{{p}_{{y}_k|s_1}}{{p}_{{y}_k|s_0}}\bigg]\\
        &\stackrel{(g)}{=}\sum_{k=1}^\infty E_j\bigg[\mathbb{I}_{T\geq k}\bigg]E_j\bigg[\log\frac{{p}_{{y}_k|s_1}}{{p}_{{y}_k|s_0}}\bigg]\\
        &\stackrel{(h)}{=}\begin{cases}
                \sum_{k=1}^\infty P[T\geq k|s_1]D({p}_{{y}_k|s_1}||{p}_{{y}_k|s_0}), \quad j=1,\\
                \sum_{k=1}^\infty -P[T\geq k|s_0]D({p}_{{y}_k|s_0}||{p}_{{y}_k|s_1}),\quad j=0,
        \end{cases}\\
        &(i)\begin{cases}
                \leq E_1[T]\sum_{k=1}^\infty\frac{D({p}_{{y}_k|s_1}||{p}_{{y}_k|s_0})}{k},\quad j=1,\\
                \geq-E_0[T]\sum_{k=1}^\infty\frac{D({p}_{{y}_k|s_0}||{p}_{{y}_k|s_1})}{k},\quad j=0,
        \end{cases}
\end{align*}
Step $(g)$ can be shown with Lemma 1. In $(h)$, the first term equivalence can be found in \cite{leon-garcia_probability_2008} and the second term equivalence is simply the definition of K-L distance. Lastly, $(i)$ is achieved using the Markov inequality\cite{leon-garcia_probability_2008}. 

Assuming that the log-likelihood ratio exceeds the bounds only slightly at the time of termination, then the following approximations can be made:
\begin{align*}
        &\!\begin{cases}
        E_1[\log(L_T(\mathbf{y}_T))]\!\approx\!\log B\leq E_1[T]\sum_{k=1}^\infty\!\frac{D({p}_{{y}_k|s_1}||{p}_{{y}_k|s_0})}{k},\\
        \!-E_0[\log(L_T(\mathbf{y}_T))]\!\approx\!-\!\log A\!\leq\! E_0[T]\!\sum_{k=1}^\infty\!\frac{D({p}_{{y}_k|s_0}||{p}_{{y}_k|s_1})}{k},\\
        \end{cases}\\
        &\Rightarrow\begin{cases}
        \frac{\log B}{E_1[T]}\leq\sum_{k=1}^\infty\frac{D({p}_{{y}_k|s_1}||{p}_{{y}_k|s_0})}{k},\\
        \frac{-\log A}{E_0[T]}\leq\sum_{k=1}^\infty\frac{D({p}_{{y}_k|s_0}||{p}_{{y}_k|s_1})}{k},\\
        \end{cases}\\
        &\!\stackrel{(j)}{\Rightarrow}\!\!\begin{cases}
        \!\!\frac{\log B}{E_1[T]}\!\leq\!\sum_{k=1}^\infty\!\frac{1}{k}\!\bigg(\!\bar\lambda_{k|1}\!(\!\mathbf{x}_1\!)\!\log\!\!\bigg(\!\!\frac{\bar\lambda_{k|1}(\mathbf{x}_1)}{\bar\lambda_{k|0}(\mathbf{x}_1)}\!\!\bigg)\!\!+\!\!\bar\lambda_{k|0}(\!\mathbf{x}_1\!)\!-\!\bar\lambda_{k|1}\!(\!\mathbf{x}_1\!)\!\!\bigg)\!,\\
        \!\!\frac{-\log A}{E_0[T]}\!\leq\!\sum_{k=1}^\infty\!\frac{1}{k}\!\bigg(\!\bar\lambda_{k|0}\!(\!\mathbf{x}_1\!)\!\log\!\!\bigg(\!\!\frac{\bar\lambda_{k|0}(\mathbf{x}_1)}{\bar\lambda_{k|1}(\mathbf{x}_1)}\!\!\bigg)\!\!+\!\!\bar\lambda_{k|1}(\!\mathbf{x}_1\!)\!-\!\bar\lambda_{k|0}\!(\!\mathbf{x}_1\!)\!\!\bigg)\!.
        \end{cases}
\end{align*}
In $(j)$, we specialise by employing the K-L distances for the key Poisson distributions. From these inequalities, it can be seen that the bounds are only non-trivial if the infinite sums are convergent. Thus, we partition the sets of indices in the sum into those less than an \emph{effective} stopping time and those larger. Then, there exists $\epsilon>0$ and $T_j(\epsilon)$, the desired expected stopping time under $s_j$ for $j\in\{0,1\}$, yielding Proposition 2:
\begin{align*}
        \!\begin{cases}
        \!\!\sum_{k=T_1(\epsilon)+1}^{\infty}\frac{1}{k}\!\bigg(\!\bar\lambda_{k|1}\!(\!\mathbf{x}_1\!)\!\log\!\!\bigg(\!\!\frac{\bar\lambda_{k|1}(\mathbf{x}_1)}{\bar\lambda_{k|0}(\mathbf{x}_1)}\!\!\bigg)\!\!+\!\!\bar\lambda_{k|0}(\!\mathbf{x}_1\!)\!-\!\bar\lambda_{k|1}\!(\!\mathbf{x}_1\!)\!\!\bigg)\!\leq\epsilon,\\
        \!\!\sum_{k=T_0(\epsilon)+1}^{\infty}\frac{1}{k}\!\bigg(\!\bar\lambda_{k|0}\!(\!\mathbf{x}_1\!)\!\log\!\!\bigg(\!\!\frac{\bar\lambda_{k|0}(\mathbf{x}_1)}{\bar\lambda_{k|1}(\mathbf{x}_1)}\!\!\bigg)\!\!+\!\!\bar\lambda_{k|1}(\!\mathbf{x}_1\!)\!-\!\bar\lambda_{k|0}\!(\!\mathbf{x}_1\!)\!\!\bigg)\!\leq\epsilon,\\
        \end{cases}
\end{align*}
and
\begin{align*}
        \!\begin{cases}
        \!\!\frac{\log B}{T_1(\epsilon)}\leq\sum_{k=1}^{T_1(\epsilon)}\frac{1}{k}\!\bigg(\!\bar\lambda_{k|1}\!(\!\mathbf{x}_1\!)\!\log\!\!\bigg(\!\!\frac{\bar\lambda_{k|1}(\mathbf{x}_1)}{\bar\lambda_{k|0}(\mathbf{x}_1)}\!\!\bigg)\!\!+\!\!\bar\lambda_{k|0}(\!\mathbf{x}_1\!)\!-\!\bar\lambda_{k|1}\!(\!\mathbf{x}_1\!)\!\!\bigg)\!,\\
        \!\!\frac{-\log A}{T_0(\epsilon)}\leq\sum_{k=1}^{T_0(\epsilon)}\frac{1}{k}\!\bigg(\!\bar\lambda_{k|0}\!(\!\mathbf{x}_1\!)\!\log\!\!\bigg(\!\!\frac{\bar\lambda_{k|0}(\mathbf{x}_1)}{\bar\lambda_{k|1}(\mathbf{x}_1)}\!\!\bigg)\!\!+\!\!\bar\lambda_{k|1}(\!\mathbf{x}_1\!)\!-\!\bar\lambda_{k|0}\!(\!\mathbf{x}_1\!)\!\!\bigg)\!.
        \end{cases}
\end{align*}

Using Propositions 1 and 2, we assume that all past transmitted source symbols are $s_1$.  This leads to maximal ISI. Our desired optimisation is given by $P1$:
\begin{align*}
        &\hat{\mathbf{x}}_1\!\!=\!\argmin_{\mathbf{x}_1}\!\bigg(\!\pi_1(\tau)\!+\!\mu\sum_{i=2}^N\pi_i(\tau)\!\bigg)\!x_{1|1}\!+\!\mu\!\sum_{k=1}^{N-1}x_{k+1|1}\!\!\sum_{i=1}^{N-k}\!\pi_i(\tau)\\
        &\mbox{s.t.}~\sum_{k=1}^{T_1(\epsilon)}\!\frac{1}{k}\!\bigg(\!\bar\lambda_{k|1}\!(\!\mathbf{x}_1\!)\!\log\!\bigg(\!\frac{\bar\lambda_{k|1}(\!\mathbf{x}_1\!)}{\bar\lambda_{k|0}\!(\!\mathbf{x}_1\!)}\!\bigg)\!+\!\bar\lambda_{k|0}\!(\!\mathbf{x}_1\!)\!-\!\bar\lambda_{k|1}\!(\!\mathbf{x}_1\!)\!\bigg)\!\!\geq\!\!\frac{\log\! B}{T_1(\epsilon)}\\
        &~~~~\sum_{k=1}^{T_0(\epsilon)}\!\frac{1}{k}\!\bigg(\!\bar\lambda_{k|0}\!(\!\mathbf{x}_1\!)\!\log\!\bigg(\!\frac{\bar\lambda_{k|0}(\!\mathbf{x}_1\!)}{\bar\lambda_{k|1}\!(\!\mathbf{x}_1\!)}\!\bigg)\!+\!\bar\lambda_{k|1}\!(\!\mathbf{x}_1\!)\!-\!\bar\lambda_{k|0}\!(\!\mathbf{x}_1\!)\!\bigg)\!\!\geq\!\!\frac{\!-\!\log\! A}{T_0(\epsilon)}\\
        &~~~~||\mathbf{x}_1||_2\leq P,
\end{align*}  
where $\bar\lambda_{k|1}(\mathbf{x}_1)=\pmb{\pi}_{k+N\cdot B}(\tau)\otimes[\mathbf{x}_1;\dots;\mathbf{x}_1;\mathbf{x}_1(1:k)]+\lambda_0t_s$ and $\bar\lambda_{k|0}(\mathbf{x}_1)=\pmb{\pi}_{k+N\cdot B}(\tau)\otimes[\mathbf{x}_1;\dots;\mathbf{x}_1;\mathbf{0}(1:k)]+\lambda_0t_s$. Here, $[\mathbf{x}_1;\dots;\mathbf{x}_1;\mathbf{x}_1(1:k)]\in\mathbb{R}^{N\cdot B+k}$ is the concatenated vector of $B$ $\mathbf{x}_1$'s from memory and the first $k$ elements of $\mathbf{x}_1$, where the notation $(1:k)$ denotes the firs $k$ elements of a vector; similarly for $[\mathbf{x}_1;\dots;\mathbf{x}_1;\mathbf{0}(1:k)]$. This can be solved using traditional optimisation algorithms such as the interior-point method. We emphasise that this optimisation process only needs to be done once offline; thus it does not contribute to transmitter complexity. 

\section{Numerical Results}
\label{sec:results}

\begin{table}[h!]
\centering
\caption{Example numerical results}
\begin{tabular}{|c|c|c|c|c|c|c|}
\hline
$t_s$ (s) & $\tau$ (s) & $\bar{T}_0$ & $\bar{T}_1$ & $T_0(\epsilon)$ & $T_1(\epsilon)$ & $||\hat{\mathbf{x}}_1||_2$ \\ \hline
0.1 & 0 & 4.7 & 4.2 & 5 & 5 & 56.1 \\ \hline
0.1 & 0.5 & 4.8 & 5.5 & 5 & 5 & 56.1 \\ \hline
0.025 & 0 & 4.8 & 4.9 & 5 & 5 & 99.6 \\ \hline
\end{tabular}
\label{tab:ex_results}
\end{table}

The following parameters are used for simulation: $\alpha=\beta=10^{-3}$, $\rho=\sqrt{0.3}$, $N=20$, $\lambda_0=4~\mbox{molecules/s}$, $T_0(\epsilon)=T_1(\epsilon)=5$ and $P=100$. The transmission rate is controlled by changing the sampling rate $t_s$. Packet lengths of $10^4$ bits are considered. We solve the optimisation problem presented in Section \ref{sec:opt}, assuming $\tau=0$, using the interior-point method in MATLAB to show the robustness of MASPRT to synchronisation errors. It is found that as the sampling rate increases, the magnitude of the vector $\hat{\mathbf{x}}_1$ increases. In the presence of synchronisation error, let $\bar{T}_0$ and $\bar{T}_1$ denote the empirical mean of the stopping times under $s_0$ and $s_1$, respectively, it is found that $\bar{T}_0$ stays roughly the same while $\bar{T}_1$ increases. Examples of simulation results are shown in Table \ref{tab:ex_results}. This follows from the decay of the hitting probabilities; as $t_s$ decreases or as $\tau$ increases, the hitting probabilities decrease. Therefore the transmitter compensates by increasing the number of molecules transmitted in the case of increasing transmission rate, or the receiver requires more samples to converge in the case of increasing synchronisation error. Since $\mathbf{x}_0=\mathbf{0}$, synchronisation errors do not affect the expected stopping time of the algorithm under $s_0$.
 

The \emph{memory-limited decision aided} (MLDA) decoder and the \emph{asynchronous detector with decision feedback} (ADDF), proposed in \cite{mosayebi_receivers_2014} and \cite{noel_asynchronous_2017}, respectively, are used as competing schemes to show that MASPRT can perform similarly to MLDA under synchronous scenarios and improve asynchronous detection performance over both schemes. The MLDA transmitter modulates the signal by mapping $s_j\in\mathbb{R}\rightarrow x_j\in\mathbb{R}$, for $j\in\{0,1\}$. For each sample $Y_i$, MLDA makes a decision using an estimate of the ISI ($\hat{\lambda}^{ISI}_i$) from its memory and the maximum likelihood estimator $\hat{x}=\max\limits_jP[Y_i=y_i|x_j,\hat{\lambda}^{ISI}_i]$. A threshold can be derived
\begin{equation*}
    \gamma=\frac{\pi_1(x_1-x_0)}{\log\big(\frac{\pi_1x_1+\hat{\lambda}^{ISI}_i+n_0}{\pi_1x_0+\hat{\lambda}^{ISI}_i+n_0}\big)}.
\end{equation*}
The detection rule is then formulated as:
\begin{eqnarray*}
\delta(y_i) & = & \left\{ \begin{array}{l l}
s_0, & y_i < \gamma,\\
s_1, &y_i \geq \gamma.
\end{array}
\right.
 \label{eq:decision_MLDA}
\end{eqnarray*}
In the paper, it is shown that this scheme is near optimal under synchronous transceivers. This is found to be true as our simulations showed MLDA to be best in terms of bit error ratio (BER) among the schemes considered here over a range of transmission rates for $\tau=0$s.

ADDF oversamples each source symbol and thresholds the maximum sample within the window of observation to determine if the transmitted source symbol is $s_0$ or $s_1$. The transmitter modulation mapping is the same as MLDA. Consider $N$ samples within a particular source symbol duration, $[y_1,y_2,\dots,y_N]$, the receiver subtracts the estimated expected ISI ($\hat{\lambda}^{ISI}_i$) using its memory, and finds the maximum sample within the window of observation:
\begin{equation*}
    y_{max}~~=\max\limits_{i\in\{1,\dots,N\}}y_i-\hat{\lambda}^{ISI}_i,
\end{equation*}
$y_{max}$ is then thresholded to decide between $s_0$ and $s_1$
\begin{eqnarray*}
\delta(y_{max}) & = & \left\{ \begin{array}{l l}
s_0, & y_{max} < \eta,\\
s_1, &y_{max} \geq \eta.
\end{array}
\right.
 \label{eq:decision_ADDF}
\end{eqnarray*}
If the synchronisation error is small, the maximum will exist within the window of observation. The decision threshold $\eta$ is optimised by calculating the probability of error for a range of $\eta$'s and choosing the value that gives the lowest error probability.



\begin{figure}[t!]
        \centering 
        \includegraphics[width=0.9\linewidth]{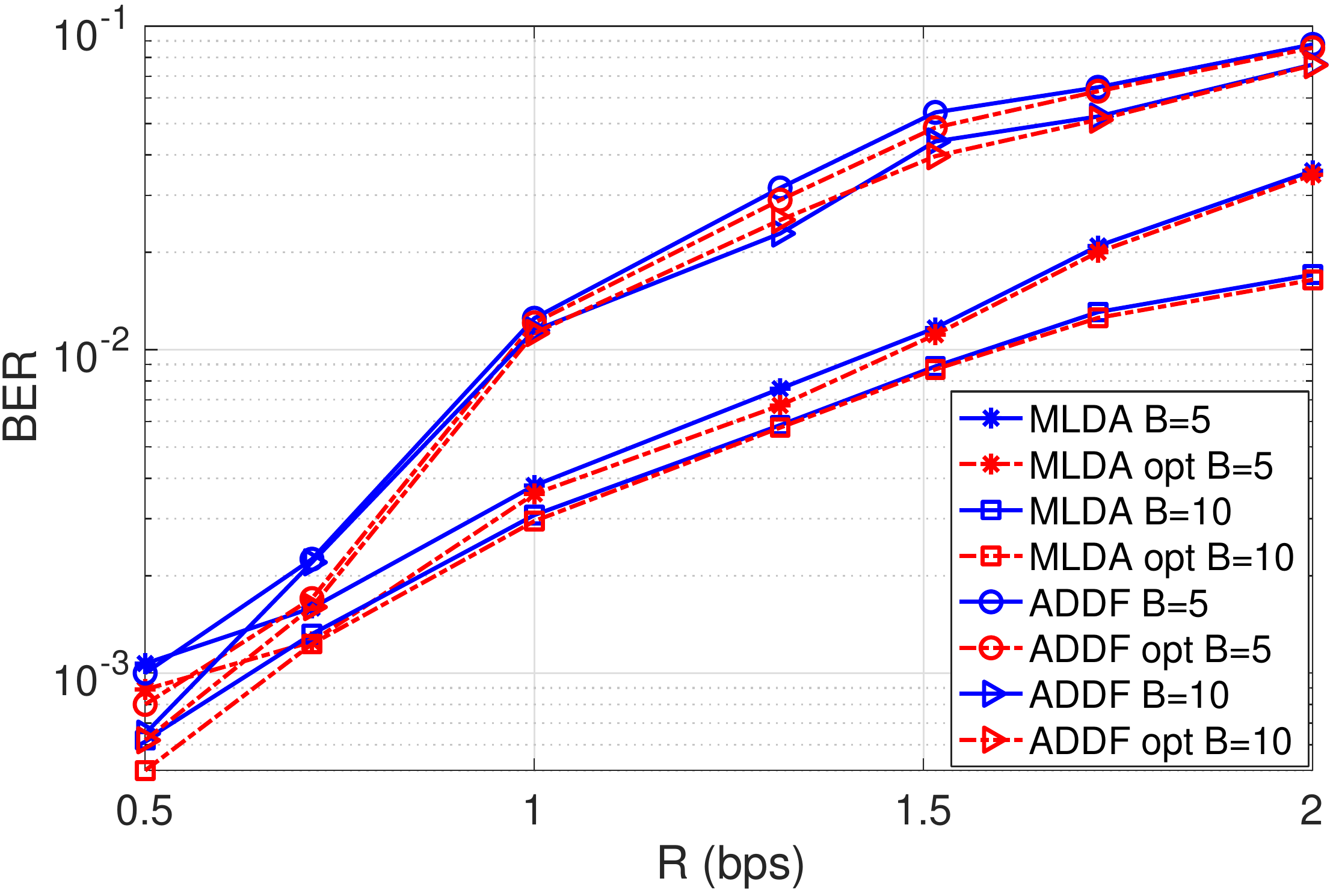}
        \vspace{-0.3cm}
        \caption{Comparison of MLDA and ADDF using their original modulation mapping and the optimised $\hat{\mathbf{x}}_1$ sequences for each transmission rate assuming $\tau=0$. $B$ is the number of bits in the receiver memory. MLDA and ADDF are simulated with $x_0=0$ and $x_1=100$. "MLDA opt" and "ADDF opt" refer to using the optimised $\hat{\mathbf{x}}_1$ sequences. Results are shown for $\tau=0.1$s.}
        \label{fig:ber_bps_opt}
        \vspace{-0.3cm}
\end{figure}

Figure \ref{fig:ber_bps_opt} compares the BER at different transmission rates for MLDA and ADDF running with their original modulation mapping and the optimised $\hat{\mathbf{x}}_1$ sequences. Here ``MLDA opt" and ``ADDF opt", refer to each scheme using the optimised $\hat{\mathbf{x}}_1$ sequences. It can be seen that using the optimised sequences provide minor improvements over the original modulation mapping, mostly apparent at low transmission rates. This is due to the optimised sequences having low magnitude at low transmission rates, reducing the effect of ISI. At high transmission rates, the results are almost identical between using the original mapping and the optimised sequences. Henceforth, the MLDA and ADDF modulation mappings will use the optimal $\hat{\mathbf{x}}_1$ sequences.

\begin{figure}[t!]
        \centering 
        \includegraphics[width=0.9\linewidth]{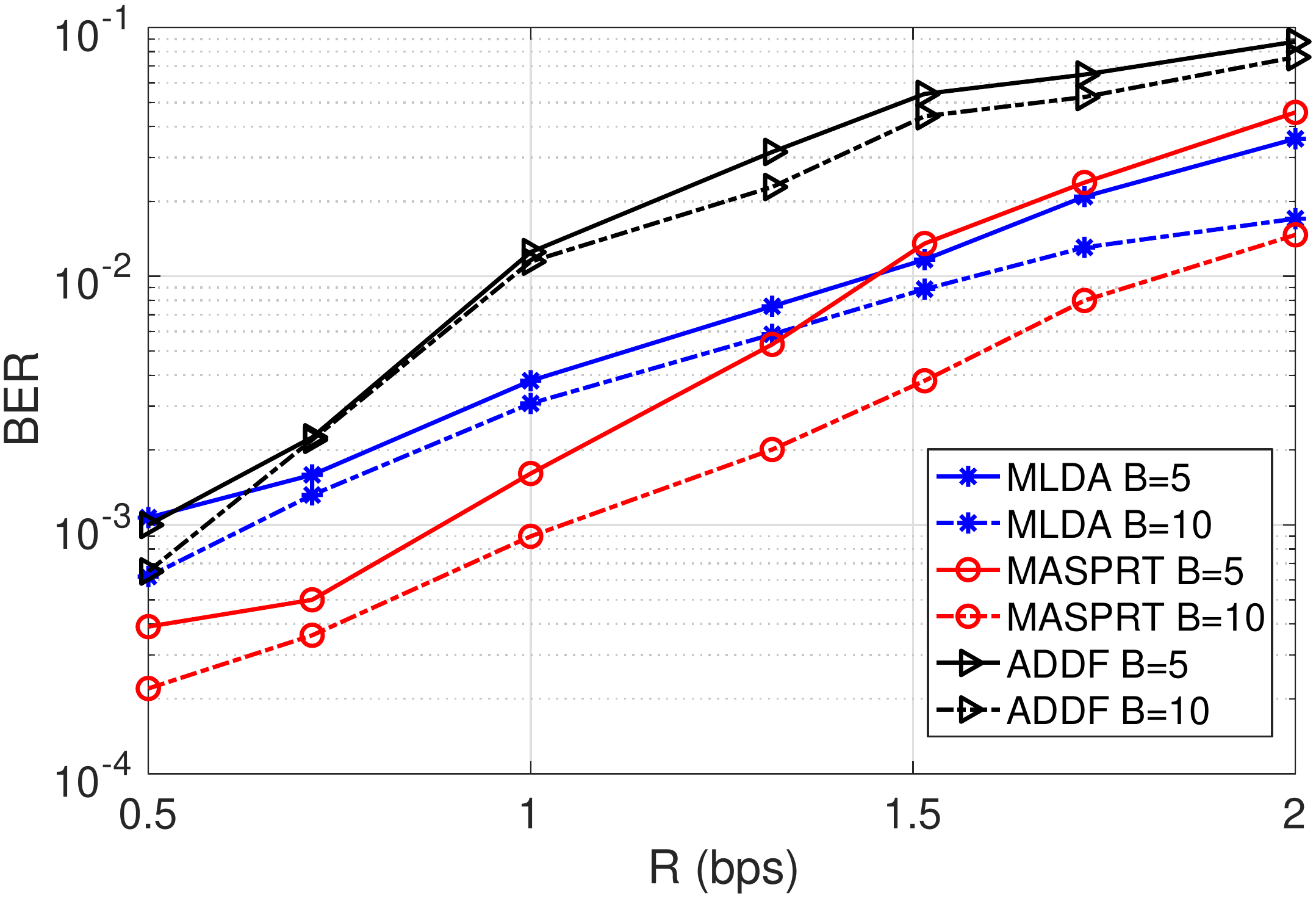}
        \vspace{-0.3cm}
        \caption{Plotting transmission rate (R) against bit error ratio (BER) for $\tau=0.1$s. All schemes here are simulated with the optimised $\hat{\mathbf{x}}_1$ sequences for each transmission rate assuming $\tau=0$.}
        \label{fig:ber_bps}
        \vspace{-0.3cm}
\end{figure}

Figure \ref{fig:ber_bps} shows the BER at different transmission rates. The diffusion rate sequence $\mathbf{x}_1$ are optimised for each transmission rate. The synchronisation error is set at $\tau=0.1$s. In all scenarios, MASPRT performs better than MLDA and ADDF in the presence of synchronisation error. MASPRT also improves significantly with increasing memory whereas MLDA improves only marginally. The ADDF scheme performs significantly worse than both MLDA and MASPRT due to the fact that it only considers the first moment of the ISI distribution.

\begin{figure}[t!]
        \centering
        \includegraphics[width=0.9\linewidth]{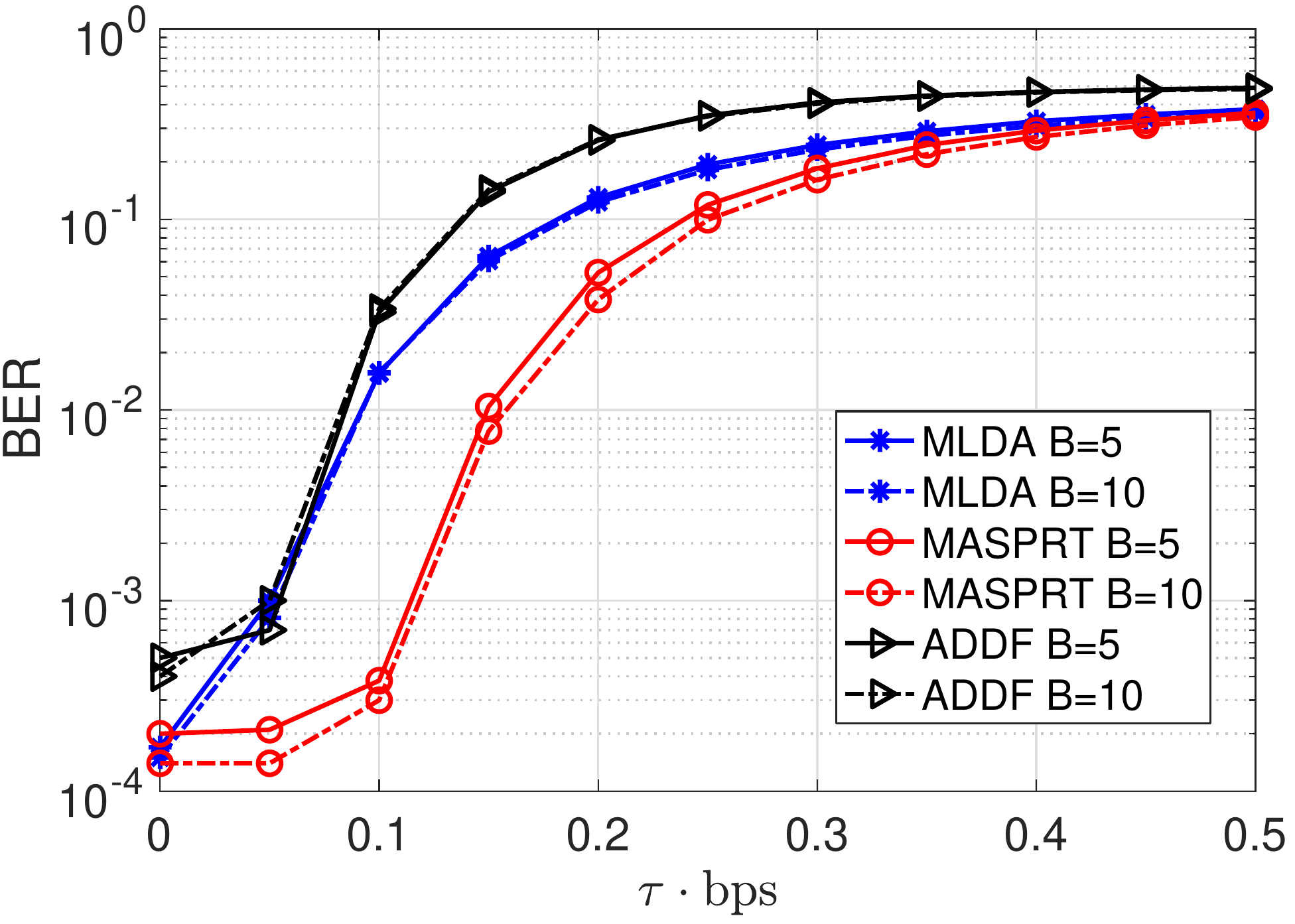}
        \vspace{-0.3cm}
        \caption{Numerical results of plotting normalised synchronisation error ($\tau\cdot\mbox{bps}$) against bit error ratio (BER). Transmission rate is set at 0.5 bps. All schemes here are simulated with the optimised $\hat{\mathbf{x}}_1$ sequence for $R=0.5$ bps.}
        \label{fig:ber_async}
        \vspace{-0.3cm}
\end{figure} 

Figure \ref{fig:ber_async} shows the resilience of each scheme to synchronisation error $\tau$. Here, $R=0.5$ bps and $\mathbf{x}_1$ is optimised for this transmission rate. The results show that MASPRT is able to maintain a roughly constant BER up to approximately $\tau=0.2$s, whereas the MLDA scheme does not have this resilience. On average, MASPRT is approximately 2 and 2.5 times better in BER than MLDA and ADDF, respectively. The performance improvement comes from the fact that MASPRT is designed to stop before the sampling limit. Thus, when a synchronisation error is present, the accumulation of errors in the calculation of LLR is lower compared to MLDA, resulting in lower error probability. Moreover, whereas MASPRT improves slightly with increased memory, MLDA does not, as both $B=5$ and $B=10$ cases tested here show almost the exact same results. ADDF can also be seen to tolerate some synchronisation errors, but it's overall higher BER means it cannot perform as well as MLDA or MASPRT.

\vspace{-0.2cm}

\section{Conclusions}
\label{sec:conclusions}

A transceiver design for molecular communication, called {\em Memory Aided Sequential Probability Ratio Test} (MASPRT), based on SPRT has been proposed in this paper. Its error probability and expected stopping time have been bounded and the bounds were used to optimise the transmitted signal design. MASPRT improves asynchronous detection performance for the same data rate as MLDA and ADDF by a reduction in bit error ratio by a factor of 1/2 and can tolerate up to 0.2s of synchronisation error.

\vspace{-0.2cm}


%



\ifCLASSOPTIONcaptionsoff
  \newpage
\fi


\bibliographystyle{IEEEtran}
\bibliography{MBMC.bib}

\end{document}